\documentclass[11pt,a4paper]{article}
\date{}
\usepackage{amsmath}
\usepackage{float}
\usepackage{graphicx}
\usepackage[english]{babel}
\usepackage{amssymb}
\usepackage{mathrsfs}

\title{The Lamb Problem with a Nonuniform String}

\author{A. Jahan\\
Research Institute for Astronomy and Astrophysics of Maragha (RIAAM)\\University of Maragheh, P. O. Box 55136-553, Maragheh, Iran\\
\textsf{jahan@maragheh.ac.ir}}

\begin{document}
\maketitle

\begin{abstract}

The time evolution of an oscillator coupled to an infinite string with a discontinuous mass density is investigated. It is shown that the equation of motion of the oscillator leads to a nonlinear characteristic equation due to the frequency-dependent nature of the point impedance of the string. Then, the method of perturbation theory is applied to solve the characteristic equation to first order in perturbation parameter. Finally, response of the oscillator to an impulse is obtained by means of the Green's function method.   \\
Keywords: Lamb model; nonuniform string; characteristic equation; perturbation theory; Green's function; point impedance.
\end{abstract}
\section{Introduction}
The Lamb system, originally introduced by H. Lamb in 1900, is a simple setup made of an oscillator coupled to an infinite taut string [1]. It is a simple and useful model to realize the \textit{radiation damping}. When the oscillator begins to oscillate, its energy diminishes by the waves emitted along the string, which in turn, results in a dissipative force on the motion of oscillator, similar to the phenomenon that takes place in classical electrodynamics when a charged oscillator loses energy by emitting the electromagnetic waves.\\
The original Lamb model is linear, i.e. the oscillator is harmonic. A nonlinear oscillator leads to chaotic reflected and transmitted waves [2]. In a model with two nonlinear oscillators, the reciprocity is lost, which means that the same wave transmits differently in two opposite directions [3]. The long time asymptotics for an anharmonic oscillator are obtained in [4] where it is shown that the finite energy solution decays to a sum of dispersive and stationary waves. A similar setup with massless case is studied in [5]. The lamb system with finite and infinite domains and dispersive medium is investigated in [6]. The problem of instability in Lamb system and its gyroscopic variant in $n$-dimensions is studied in [7]. The scattering problem in the Lamb model is presented by some textbooks, where the equation of motion of the oscillator, i.e. equation (13) is employed to achieve the scattering coefficients [8, 9]. \\
In this work, we consider a Lamb model with an infinite string made of two semi-finite strings of different mass densities $\rho_{-}$ and $\rho_{+}$ joined together at $x=0$. An interesting feature of such a string is that it has a frequency dependent point impedance. So, since the impedance of the string appears in the equation governing the motion of oscillator, it leads to a nonlinear characteristic equation. Therefore, we face with a problem which is of interest from the perspective of the classical perturbation theory.  \\
This paper consists of 6 sections. The section 2, is a review of the problem and the mathematical apparatus used throughout the paper. An emphasis is put on the use of Green's function because the Green's function of a free string yields its impedance. Also, response of the oscillator to an impulse can be obtained using the Green's function of the whole system, i.e. the string plus the oscillator. However, one should be able to analyze the problem solely by using the wave function. In section 3, the equation of motion of the oscillator is obtained in terms of the impedance of the string, starting from the Lagrangian density of the problem. Since the string has a discontinuous mass density, its impedance is a frequency-dependent quantity which in turn leads to a nonlinear characteristic equation for the oscillator. So, in section 4 we apply the perturbation theory to obtain the damping factor and oscillation frequency of the oscillator to first order in expansion parameter. In section 5, we obtain the response of oscillator to an impulse by evaluating the Green's function of the system. We find that the oscillator executes a damped motion with an amplitude decaying exponentially in time. Finally, a short conclusion is included in section 6.
\section{The Lamb problem}
We consider an infinite taut string with a constant tension $T$ and a nonuniform mass density $\mu(x)$. The field $\phi(x,t)$ denotes the displacement of string from the equilibrium. The Lagrangian density for the displacement field is given by
\begin{equation}\label{3}
\mathcal L=\frac{1}{2}\mu(x)\bigg(\frac{\partial\phi(x,t)}{\partial t}\bigg)^2-\frac{1}{2}T\bigg(\frac{\partial\phi(x,t)}{\partial x}\bigg)^2+\phi(x,t)F(x,t)+\mathcal L_{int}.
\end{equation}
\begin{figure}[h!]
\centering
\includegraphics[width=85mm]{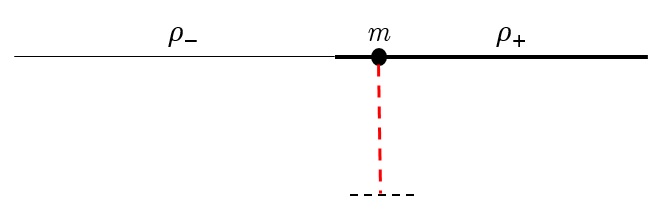}
\caption{An infinite string made of two different semi-infinite strings connected at $x=0$ . An oscillator (dashed line) is coupled to the string at $x_0$.}
\end{figure}
Here $F(x,t)$ is an external force acting on the string and $\mathcal{L}_{int}$ denotes the interaction of string with its environment. Inserting the above Lagrangian into the Euler-Lagrange equation yields the equation of motion
\begin{equation}\label{1}
\mu(x)\frac{\partial^2\phi(x,t)}{\partial t^2}-T\frac{\partial^2\phi(x,t)}{\partial x^2}-\frac{\partial\mathcal L_{int}}{\partial\phi}=F(x,t).
\end{equation}
For a harmonic oscillator with mass $m$ attached to string at $x_0$, the mass density and interaction Lagrangian can be written
{\setlength\arraycolsep{2pt}
\begin{eqnarray}\label{1}
\mu(x) &=&\rho(x)+m\delta(x-x_0),\\
\mathcal L_{int}&=&-\frac{1}{2}\kappa\phi^2(x,t)\delta(x-x_0).
\end{eqnarray}}
where $\kappa=m\alpha^2$ is stiffness of the spring and $\rho(x)$ is the mass density of the free string. Thus, the relations (3) and (4) modify (2) to
\begin{equation}\label{1}
G^{-1}\phi(x,t)=\frac{1}{T}F(x,t),
\end{equation}
with
\begin{equation}\label{1}
G^{-1}=\left(\frac{1}{v^2}\frac{\partial^2}{\partial t^2}-\frac{\partial^2}{\partial x^2}\right)+\frac{1}{T}\delta(x-x_0)\left(m\frac{\partial^2}{\partial t^2}+\kappa^2\right).
\end{equation}
where the speed of wave is denoted by $v=\big(\frac{T}{\rho}\big)^{\frac{1}{2}}$. A formal solution of equation (5) is given by
\begin{eqnarray}\label{1}
\phi(x,t)=\frac{1}{T}\int^\infty_{-\infty}dx'\int^\infty_{-\infty}dt'G(xt;x't')F(x',t'),\qquad\qquad\qquad t'<t.
\end{eqnarray}
where $G(xt;x't')$ is the retarded Green's function of the system satisfying the relation
\begin{eqnarray}\label{1}
G^{-1}G(xt;x't')=\delta(t-t')\delta(x-x'),
\end{eqnarray}
and can be written
\begin{equation}\label{1}
G(xt;x't')
=\int_{-\infty}^\infty \frac{d\omega}{2\pi}e^{-i\omega(t-t')}g(x,x';\omega).
\end{equation}
with $g(x,x';\omega)$ as its Fourier transform. Then, inserting (9) into (8) reveals that
\begin{eqnarray}\label{1}
\left(\frac{\partial^2}{\partial x^2}+k^2(x)+V(x)\right)g(x,x';\omega)=-\delta(x-x'),\qquad\qquad\qquad k(x)=\frac{\omega}{v(x)},
\end{eqnarray}
where the potential term is
\begin{equation}\label{1}
V(x)=\frac{m}{T}(\omega^2-\alpha^2)\delta(x).
\end{equation}
In absence of the oscillator, the Green's function $g(x,x';\omega)$ reduces to the Green's function of the free string $g_0(x,x';\omega)$ for which we have
\begin{eqnarray}\label{1}
\left(\frac{\partial^2}{\partial x^2}+k^2(x)\right)g_0(x,x';\omega)=-\delta(x-x').
\end{eqnarray}

\section{Equation of motion of the oscillator}
The back-reaction of string on the motion of oscillator arises as a dissipative term depending on the impedance of string. To see how this happens, let us first set $F=0$ in (5) and then integrate both sides over an infinitesimal interval including the oscillator. So, we obtain [1-9]
\begin{eqnarray}\label{1}
\left(m\frac{d^2}{dt^2}+\kappa \right)\xi(t)=T\frac{\partial\phi(x,t)}{\partial x}\bigg|^{x_0+\epsilon}_{x_0-\epsilon}.
\end{eqnarray}
where $\xi(t)\equiv\phi(x_0,t)$ denotes the displacement of oscillator. Then with the aid of equation (A.16) we arrive at
\begin{eqnarray}\label{1}
\ddot{\xi}+\frac{1}{m}Z(\omega,x_0)\dot{\xi}+\alpha^2\xi=0.
\end{eqnarray}
which exhibits the role of impedance of the string on the dynamics of oscillator. Assuming a time harmonic solution of the form $\xi\sim e^{-i\omega t}$ yields a nonlinear characteristic equation of the form
\begin{equation}\label{1}
\omega^2+\frac{i}{m}\omega Z(\omega,x_0)-\alpha^2=0.
\end{equation}
This equation poses two solutions of the form
{\setlength\arraycolsep{2pt}
\begin{eqnarray}\label{1}
\omega&=&h_{+}(\omega),\\
\omega&=&h_{-}(\omega),
\end{eqnarray}}
where the complex-valued eigen-frequencies are given by
\begin{equation}\label{1}
h_\pm (\omega)=-\frac{i}{2m}Z(\omega,x_0)\pm\sqrt{\alpha^2-\frac{Z^2(\omega,x_0)}{4m^2}}.
\end{equation}
The solutions (16) and (17) are exact only when the point impedance is a constant and independent of $\omega$. Otherwise, one must apply an approximation method to obtain the solutions of (16) and (17). Hence, the characteristic equation admits a general solution of the form
\begin{equation}\label{1}
\xi(t)=a_+e^{ih_+ t}+a_-e^{ih_- t}.
\end{equation}
For an under-damped oscillator with ${\mathcal{Z}_-}<2m\alpha$ and the initial condition $\xi(0)=0$, the equation (19) yields the time evolution of the form
\begin{equation}\label{1}
\xi(t)=\xi_0 e^{-\gamma t}\sin\Omega t,
\end{equation}
provided that (18) takes the form
\begin{equation}\label{1}
h_\pm (\omega)=-i\gamma\pm\Omega.
\end{equation}
In next section, we obtain the damping factor $\gamma$ and oscillation frequency $\Omega$ using the method of perturbation theory. At the point $x_0=0$, the impedance is a constant (See Appendix). Thus, for an oscillator coupled at $x_0=0$, with the aid of the equations (16)-(19) and (A.19) we have
\begin{equation}\label{1}
\xi(t)=\xi_0e^{-\frac{1}{2m}(\rho_+v_++\rho_-v_-)t}\sin\left(t\sqrt{\alpha^2-\frac{(\rho_+v_++\rho_-v_-)^2}{4m^2}}\right).
\end{equation}
Also, for a uniform string with constant impedance $\mathcal Z=2\rho v$ we obtain
\begin{equation}\label{1}
\xi(t)=\xi_0 e^{-\frac{1}{m}\rho vt}\sin\left(t\sqrt{\alpha^2-\frac{\rho^2v^2}{m^2}}\right).
\end{equation}
which also can be obtained by letting $\rho_-=\rho_+=\rho$ in (22). We skip the cases $2m\alpha={Z(0)}$ and $2m\alpha<{Z(0)}$ which lead to the critically damped and over-damped motions, respectively.
\section{Perturbative solution of characteristic equation}
For an oscillator coupled at $x_0\neq0$, the impedance is frequency-dependent due to the partial reflection of the wave at the joining point.
A perturbative solution of (16) and (17) is possible by supposing an expansion of the form
\begin{equation}
\omega=\sum_{n=0}^\infty A^n_R\omega_{n,\pm}.
\end{equation}
for the both sides of (16) and (17) in terms of the reflection coefficient $A_R$ as the perturbation parameter. The reflection coefficient is a dimensionless quantity. Furthermore, the uniform string limit, i.e. $\rho_-=\rho_+$ is obtained when $A_R=0$. For the oscillator coupled at $x_0<0$, according to (A.19) the impedance is found to be
\begin{equation}\label{1}
Z(\omega,x_0)=\frac{\mathcal{Z}_- }{1+A_Re^{-i\chi\omega}},\qquad \qquad\qquad \chi=\frac{2x_0}{v_-}.
\end{equation}
Hence, provided that $A_R|e^{-i\chi\omega}|\ll1$, we can write
{\setlength\arraycolsep{2pt}
\begin{eqnarray}\label{1}
Z(\omega,x_0)\big|_{\omega=\omega_{0,+}+A_R\omega_{1,+}}&=&\mathcal{Z}_- -A_R\mathcal{Z}_-e^{-i\chi\omega_{0,+}}+O(A_R^2),\\
Z(\omega,x_0)\big|_{\omega=\omega_{0,-}+A_R\omega_{1,-}}&=&\mathcal{Z}_- -A_R\mathcal{Z}_-e^{-i\chi\omega_{0,-}}+O(A_R^2).
\end{eqnarray}}
From (24) and by expanding the both sides of (16) and (17) to first order in $A_R$, we have
{\setlength\arraycolsep{2pt}
\begin{eqnarray}
\omega_{0,+}+A_R\omega_{1,+}&=&h_+(\omega)\big|_{\omega=\omega_{0,+}+A_R\omega_{1,+}},\\
\omega_{0,-}+A_R\omega_{1,-}&=&h_-(\omega)\big|_{\omega=\omega_{0,-}+A_R\omega_{1,-}}.
\end{eqnarray}}
For the right-hand sides we obtain
{\setlength\arraycolsep{2pt}
\begin{eqnarray}
h_+(\omega)\big|_{\omega=\omega_{0,+}+A_R\omega_{1,+}}&=&-i\frac{\mathcal{Z}_-}{2m}+\sqrt{\alpha^2-\frac{\mathcal{Z}_-^2}{4m^2}}+
iA_R\frac{\mathcal{Z}_-}{2m}\frac{\omega_{0,+}}{\sqrt{\alpha^2-\frac{\mathcal{Z}_-^2}{4m^2}}}e^{-i\chi\omega_{0,+}},\\
h_-(\omega)\big|_{\omega=\omega_{0,-}+A_R\omega_{1,-}}&=&-i\frac{\mathcal{Z}_-}{2m}-\sqrt{\alpha^2-\frac{\mathcal{Z}_-^2}{4m^2}}-
iA_R\frac{\mathcal{Z}_-}{2m}\frac{\omega_{0,-}}{\sqrt{\alpha^2-\frac{\mathcal{Z}_-^2}{4m^2}}}e^{-i\chi\omega_{0,-}},
\end{eqnarray}}
which after equating the both sides in (28) and (29), leads to
{\setlength\arraycolsep{2pt}
\begin{eqnarray}
\omega_{0,\pm}&=&-i\frac{\mathcal{Z}_-}{2m}\pm\sqrt{\alpha^2-\frac{\mathcal{Z}_-^2}{4m^2}},\\
\omega_{1,\pm}&=&\pm i\frac{\mathcal{Z}_-}{2m}\frac{\omega_{0,\pm}}{\sqrt{\alpha^2-\frac{\mathcal{Z}_-^2}{4m^2}}}e^{-i\chi\omega_{0,\pm}}.
\end{eqnarray}}
But from (21), to first order in $A_R$ we should have
{\setlength\arraycolsep{2pt}
\begin{eqnarray}
\gamma&=&\gamma_0+A_R\gamma_1, \\
 \Omega&=&\Omega_0+A_R\Omega_1.
\end{eqnarray}}
Thus, with the help of (32) and (33) we get
{\setlength\arraycolsep{2pt}
\begin{eqnarray}
\gamma_1&=&\gamma_0e^{-\chi\gamma_0}\left(\frac{\gamma_0}{\Omega_0}\sin\chi\Omega_0-\cos\chi\Omega_0\right),\\
\Omega_1&=&\gamma_0e^{-\chi\gamma_0}\left(\frac{\gamma_0}{\Omega_0}\cos\chi\Omega_0+\sin\chi\Omega_0\right),
\end{eqnarray}}
and
{\setlength\arraycolsep{2pt}
\begin{eqnarray}
\gamma_0&=&\frac{\mathcal{Z}_-}{2m},\\
\Omega_0&=&\sqrt{\alpha^2-\frac{\mathcal{Z}_-^2}{4m^2}}.
\end{eqnarray}}
Then, plugging these results into (21) gives
{\setlength\arraycolsep{2pt}
\begin{eqnarray}
\gamma&=&\gamma_0+A_R\gamma_0e^{-\chi\gamma_0}\left(\frac{\gamma_0}{\Omega_0}\sin\chi\Omega_0-\cos\chi\Omega_0\right), \\
\Omega&=&\Omega_0+A_R\gamma_0e^{-\chi\gamma_0}\left(\frac{\gamma_0}{\Omega_0}\cos\chi\Omega_0+\sin\chi\Omega_0\right).
\end{eqnarray}}
Therefore, the time evolution of the oscillator to first order in $A_R$ is given by equation (20) with $\gamma$ and $\Omega$ given by the relations (40) and (41). We notice that due to the condition $A_Re^{-\chi\gamma_0}\ll1$ in (40) we have $0<\gamma$, regardless to the sign of $A_R$. Thus, the amplitude of oscillation dies out exponentially according to (20). Needless to say, the validity of first order perturbation holds only for an almost uniform string with $|A_R|\ll1$ or $\rho_-\approx\rho_+$, indeed. Otherwise, the higher order terms in (24) must be taken into account.
\section{Response to an impulse}
Further insight about the motion of oscillator could be obtained by looking at the response of oscillator to an impulse. The response to sudden force $F(x,t)=J_0\delta(x-x_0)\delta(t)$ delivered to the oscillator is attainable from (7) via
{\setlength\arraycolsep{2pt}
\begin{eqnarray}\label{1}
\nonumber
\phi(x_0,t)&=&\frac{J_0}{T}G(x_0t;x_00)\\
&=&\frac{J_0}{T}\int_{-\infty}^\infty \frac{d\omega}{2\pi}e^{-i\omega t}g(x_0,x_0;\omega).
\end{eqnarray}}
since $\xi(t)=\phi(x_0,t)$. On the other hand, the equation (10) admits the following solution for the Green's function
{\setlength\arraycolsep{2pt}
\begin{eqnarray}\label{1}
\nonumber
g(x,x';\omega)&=&g_0(x,x';\omega)+\int^\infty_{-\infty}dx^{\prime\prime}g_0(x,x^{\prime\prime};\omega)V(x^{\prime\prime})g(x^{\prime\prime},x';\omega)\\
&=&g_0(x,x';\omega)+\frac{m}{T}(\omega^2-\alpha^2)g_0(x,x_0;\omega)g(x_0,x';\omega).
\end{eqnarray}}
From this expression one can obtain $g(x_0,x';\omega)$ by letting $x=x_0$, on the both sides of (43). Substituting the resultant term back into (43), then yields
{\setlength\arraycolsep{2pt}
\begin{eqnarray}\label{1}
\nonumber
g(x,x_0;\omega)&=&\frac{g_0(x,x_0;\omega)}{1-\frac{m}{T}(\omega^2-\alpha^2) g_0(x_0,x_0;\omega)}\\
&=&-\frac{T}{m}\frac{1}{\omega^2+\frac{i}{m}Z(\omega,x_0)\omega-\alpha^2}\;\frac{g_0(x,x_0;\omega)}{g_0(x_0,x_0;\omega)},
\end{eqnarray}}
giving rise to
\begin{equation}\label{1}
g(x_0,x_0;\omega)=-\frac{T}{m}\frac{1}{\omega^2+\frac{i}{m}Z(\omega,x_0)\omega-\alpha^2}.
\end{equation}
This is indeed the Green's function of the oscillator described by the equation of motion, although we obtained it without refereing to (14). The denominator of this expression is the non-linear characteristic equation we encountered in (15). So, when (45) is inserted back into (42) the integrand becomes singular at the points given by (16) and (17) on the complex plane. On the other hand, the causality condition requires that the above integral to be vanished for $t<0$. This is possible only when the singular points are placed on the lower half plane, thus demanding $0<\gamma$ in (21). Therefore, upon employing the residue theorem and enclosing the integration contour in the lower (upper) half-plane for $0<t$ ($t<0$), the integral is found to be
\begin{eqnarray}\label{1}
\int_{-\infty}^\infty \frac{d\omega}{2\pi}\frac{e^{-i\omega t}}{\omega^2+\frac{i}{m}Z(\omega,x_0)\omega-\alpha^2}=-\frac{i}{h_+-h_-}\left(e^{-ih_+t}-e^{-ih_-t}\right).
\end{eqnarray}
which, after inserting into (42) and invoking (21), yields
\begin{equation}\label{1}
\phi(x_0,t)=\frac{J_0}{m\Omega} e^{-\gamma t}\sin\Omega t.
\end{equation}
This is the same result we obtained in section 3 with the amplitude $A=\frac{J_0}{m\Omega}$ and the parameters given in (40) and (41).
\section{Conclusion}
In this paper we considered a Lamb model with a string of discontinuous mass density. A discontinuous density results in a nonlinear characteristic equation for the oscillator coupled to the string. We solved the characteristic equation by applying the perturbation method and showed that to first order approximation, the oscillator evolves in time as a damped oscillator. Then using the Green's function method, we obtained the time evolution of the oscillator subjected to a sudden impulse. We found that the oscillator begins to oscillate with a decaying amplitude due to the emission of its energy along the string albeit the partial reflection of the waves at the discontinuity point.   \\

\newpage
\appendix
\numberwithin{equation}{section}
\section{Input impedance in terms of Green's function }
An infinite string made of two semi-infinite strings joined together at $x=0$. The semi-finite strings have the mass densities $\rho_{\pm}$. Thus, the mass density of string can be written as
\begin{equation}\label{1}
\rho(x)=\rho_-\Theta(-x)+\rho_+\Theta(x),
\end{equation}
which, by virtue of $k\sim\rho$, gives rise to
\begin{equation}\label{1}
k^2(x)=k^2_-\Theta(-x)+k^2_+\Theta(x),\qquad\qquad\qquad k_\pm=\frac{\omega}{v_\pm}.
\end{equation}
where $\Theta(x)$ is the Heaviside step function. Therefore, from (12) we obtain
\begin{equation}\label{1}
\left(\frac{\partial^2}{\partial x^2}+k^2_-\Theta(-x)+k^2_+\Theta(x)\right)g_0(x,x';k)=-\delta(x-x').
\end{equation}
The form of Green's function depends on the position of source. Therefore, we write
\begin{equation}\label{1}
{g_0}(x,x';k)=\mathfrak{g}_-(x,x';k)\Theta(-x')+\mathfrak{g}_+(x,x';k)\Theta(x').
\end{equation}
where $\mathfrak{g}_-$ and $\mathfrak{g}_+$ denote the corresponding Green's functions for a source located in negative ($x'<0$) or positive ($0<x'$) domains of $X$-axis. Now, to construct the Green's functions $\mathfrak{g}_{\pm}$ we use the well-know formula [10]
\begin{eqnarray}\label{1}
\mathfrak{g}(x,x';k)=\frac{\varphi_>(x)\varphi_<(x')}{W(x)}\Theta(x-x')+\frac{\varphi_>(x')\varphi_<(x)}{W(x)}\Theta(x'-x),
\end{eqnarray}
where the Wronskian is defined to be
\begin{eqnarray}\label{1}
W(x)=\varphi_>(x)\frac{d\varphi_<(x)}{dx} -\varphi_<(x)\frac{d\varphi_>(x)}{dx}.
\end{eqnarray}
The functions $\varphi_<(x)$ and $\varphi_>(x)$ are linearly independent solutions of the corresponding wave equation. For equation (A.3) these solutions are
\begin{equation}\label{1}
\varphi_<(x)=e^{-ik_-x}+A_Re^{-ik_-x},\qquad\qquad\qquad x<x'<0,\\
\end{equation}
and
\begin{equation}\label{1}
\varphi_>(x)=\left\{ \begin{array}{ll}
e^{ik_-x}+A_Re^{-ik_-x},\;\;\qquad\qquad x'<x<0,\\
A_Te^{ik_+x},\;\;\qquad\qquad\qquad\qquad x'<0<x,
\end{array} \right.
\end{equation}
with corresponding Wronskian given by
\begin{eqnarray}\label{1}
W=-{2ik_-}(1+A_R).
\end{eqnarray}
Then, the continuity of Green's function ${g_0}(x,x';k)$ and its first derivative at $x=0$, yields
{\setlength\arraycolsep{2pt}
\begin{eqnarray}\label{1}
\nonumber
A_R&=&\frac{k_--k_+}{k_++k_-}\\
&=&\frac{\mathcal{Z}_--\mathcal{Z}_+}{\mathcal{Z}_++\mathcal{Z}_-},
\end{eqnarray}}
and
{\setlength\arraycolsep{2pt}
\begin{eqnarray}\label{1}
\nonumber
A_T&=&\frac{2k_-}{k_++k_-}\\
&=&\frac{2\mathcal{Z}_-}{\mathcal{Z}_++\mathcal{Z}_-},
\end{eqnarray}}
where
{\setlength\arraycolsep{2pt}
\begin{eqnarray}\label{1}
\mathcal Z_{+}&=&\frac{2T}{v_+}=2\rho_+v_+,\\
\mathcal Z_{-}&=&\frac{2T}{v_-}=2\rho_-v_-,
\end{eqnarray}}
denote the impedances of infinite strings with mass densities $\rho_+$ and $\rho_-$. Hence, by inserting (A.7), (A.8) and (A.10) into (A.5) we arrive at the expression
\begin{equation}\label{1}
\mathfrak{g}_-(x,x';k)=\frac{i}{2k_-}\left(e^{ik_-|x-x'|}+\frac{k_--k_+}{k_++k_-}e^{-ik_-(x+x')}\right)\Theta(-x)+\left(\frac{i}{k_-+k_+}e^{ik_+x-ik_-x'}\right)\Theta(x).
\end{equation}
Similarly, by putting the source in region $0<x'$ we obtain
\begin{equation}\label{1}
\mathfrak{g}_+(x,x';k)=\left(\frac{i}{k_-+k_+}e^{ik_+x'-ik_-x}\right)\Theta(-x)+\frac{i}{2k_+}\left(e^{ik_+|x-x'|}-\frac{k_--k_+}{k_++k_-}e^{ik_+(x+x')}\right)\Theta(x).
\end{equation}
So, the Green's function of the string is given by the set of relations (A.4), (A.14) and (A.15). The point impedance of a string is defined as the ratio of the force $f$ applied at point $x_0$, to the transverse speed caused by the force, namely
{\setlength\arraycolsep{2pt}
\begin{eqnarray}\label{1}
\nonumber
Z(\omega,x_0)&=&\frac{f}{\partial_t\phi(x_0,t)}\\
&=&-\frac{T}{\partial_t\phi(x_0,t)}\frac{\partial\phi(x,t)}{\partial x}\bigg|^{x_0+\epsilon}_{x_0-\epsilon}.
\end{eqnarray}}
The above expression can be re-arranged to give
\begin{equation}\label{1}
Z(\omega,x_0)=\frac{T}{i\omega}\frac{\varphi_<(x_0)\partial_x\varphi_>(x_0)-\varphi_>(x_0)\partial_x\varphi_<(x_0)}{\varphi_<(x_0)\varphi_>(x_0)}.
\end{equation}
since $\varphi_<(x_0)=\varphi_>(x_0)$. A comparison with the formula (A.5) then yields
\begin{eqnarray}\label{1}
Z(\omega,x_0)=-\frac{T}{i\omega g_0(x_0,x_0;\omega)}.
\end{eqnarray}
Accordingly, from (A.18) we obtain the impedance
\begin{eqnarray}\label{1}
Z(\omega,x_0)=\frac{\mathcal Z_-}{1+A_Re^{-2ix_0\omega/v_-}}\Theta(-x_0)+\frac{\mathcal Z_+}{1-A_Re^{2ix_0\omega/v_+}}\Theta(x_0).
\end{eqnarray}
From the above formula, the impedance at $x_0=0$ is found to be
\begin{eqnarray}\label{1}
Z(0)=\frac{1}{2}(\mathcal Z_-+\mathcal Z_+).
\end{eqnarray}
which is just the sum of impedances of two semi-infinite strings attached together at $x_0=0$.

\end{document}